# LIGHT POLLUTION AS PART OF THE ENVIRONMENTAL PROBLEMS (2006)

Quaranta, Nancy[2], Cionco, Rodolfo Gustavo[1],

*Grupo de Estudios Ambientales-GEA*
Facultad Regional San Nicolás. Universidad Tecnológica Nacional.
Colón 332. San Nicolás. Argentina.
Email: gcionco@frsn.utn.edu.ar

[1]Facultad de Cs. Astronómicas y Geofísicas, UNLP (FCAGLP) - Instituto de Astrofísica de La Plata (CONICET)

[2]Investigador, Comisión de Investigaciones Científicas de la Provincia de Buenos Aires (CICPBA).

**Abstract**

Unscrupulous outdoor lighting produces a number of effects that are currently included under the term light pollution. Its consequences (e.g. loss of resources by energy waste), are being recognized for some time, as well as the possibility to mitigate this pollution. In the present work, we present some lines of action developed at the Facultad Regional San Nicolás of National Technological University (UTN) of Argentina to include the CL as a regular topic of study in the problems of air pollution.

## *1- Introducción y Objetivo*

La manifestación más evidente del fenómeno de la luz como agente contaminante, es el *aumento del brillo del cielo nocturno* (**ABC**). Una cantidad residual de energía no utilizada para iluminar, es enviada en forma directa o refleja a la atmósfera. Allí, es dispersada por diversas capas de aire; el cielo se "emblanquece" adquiriendo un resplandor nocturno característico ("urban sky glow", ver por ejemplo, Crawford, 2000). Consecuentemente, los objetos celestes "tienden a desaparecer" (Treanor, 1970; Walker, 1973; Gargstang, 1991; Cinzano et al., 2001). El fenómeno perjudica tan fuertemente a la astronomía, que desde hace unos 30 años trata de difundirse ampliamente junto con directivas tendientes a su mitigación. Para ejemplificar cómo puede percibirse el envío no deseado de luz al medioambiente, se muestra en la Fig.1 (panel *a)*, una zona del cielo austral observada desde las cercanías de una urbe importante, exhibiendo un alto grado de **CL**; el panel (*b*), por el contrario, muestra la misma zona pero como se observaría desde un paraje alejado de zonas urbanas.

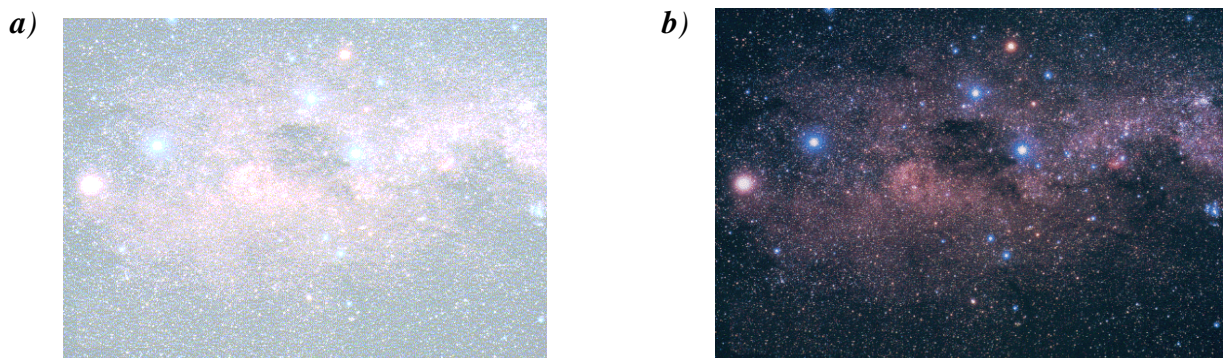

*Fig.1, (a): atmósfera con alto grado de CL; (b): la misma zona pero sin CL.*

Esta definición clásica de **CL** o *contaminación en sentido astronómico* (esto es, el **ABC**), ha sido fundamental para la toma de conciencia de este problemática como hecho *zonal y global*; aunque la luminotecnia no es ajena a los efectos *no deseados* de la iluminación, sus estándares han estado centrados en la optimización de una o de un conjunto de luminarias, más que en tratamientos zonales o globales y su impacto en el medioambiente. Hoy en día, la **CL** debe ser vista como un problema que englobe no solamente al **ABC** sino también a los efectos colaterales de la iluminación artificial tales como *intrusión lumínica, deslumbramiento* y *derroche de energía.* Esto puede explicarse de la siguiente manera: en su viaje hacia capas más altas de la atmósfera, la luz excesiva o colateral, interactúa con el medioambiente circundante, por lo tanto **CL** *en un sentido amplio*, debe referirse también a los efectos antes mencionados e incluir al **ABC**, siendo éste, el *parámetro de diagnóstico más adecuado*, ya que depende directamente de todo el conjunto de luz "exterior" mal aprovechado en un sitio o región.

La **CL** en un lugar no es un efecto puntual, sino que depende de la contribución de cada luminaria y luces de sitios cercanos; el problema es mundial, y el estado actual del mismo a escala global puede verse en la Fig.2, a través del primer *atlas mundial* de **CL** según Cinzano et al. (2001).

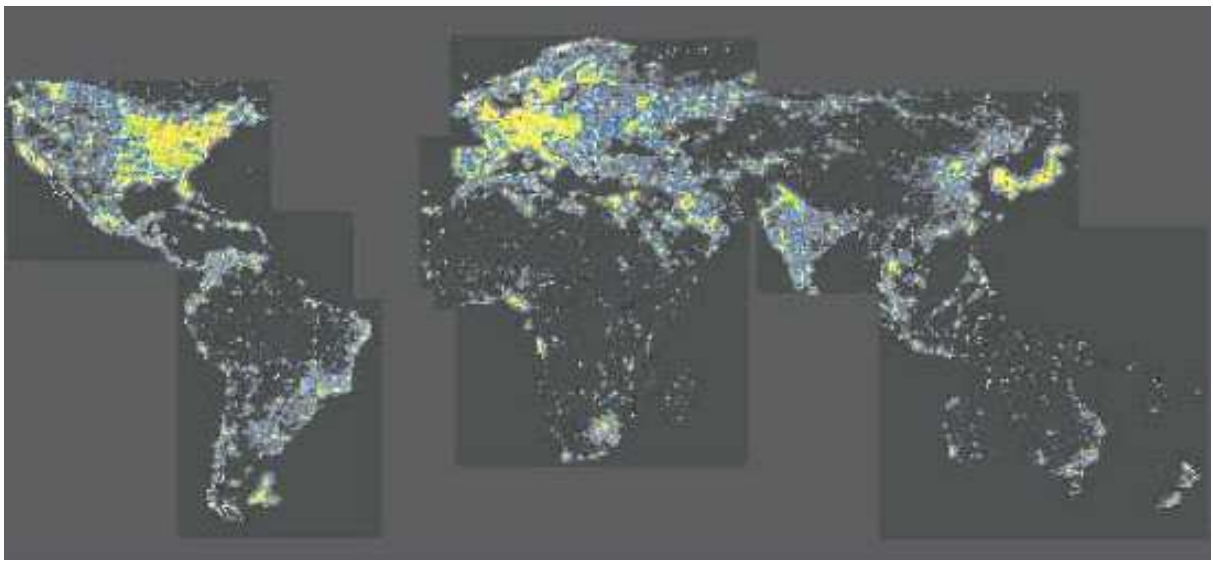

*Fig.2. Brillo zenital del cielo (banda fotométrica V) a escala mundial, medido con satélites DMSP según Cinzano et al.* Se hace evidente la asimetría Norte-Sur en el derroche de energía.

La **CL** ha comenzado a tratarse dentro del Grupo de Estudios Ambientales (GEA) de la Facultad Regional San Nicolás de la Universidad Tecnológica Nacional de Argentina (FRSN -UTN) y en el presente trabajo se exponen las líneas de acción desarrolladas en él para incluirla como tópico académico regular dentro de las problemáticas de contaminación atmosférica.

## *2 - Metodología y líneas de acción*

Se ha delineado una estrategia para difundir el tema **CL** e incluirlo en un proyecto científico-tecnológico de estudio y mitigación de problemáticas ambientales, el cual se halla en gestación en la **UTN**. **CL** comenzó a tratarse hace unos dos años en el GEA de la FRSN-UTN y básicamente las líneas de trabajo se han centrado en:

- Modelado del **ABC** de un sitio.
- Mediciones astronómicas para contrastar modelos y datos de iluminación pública.

Algunos resultados obtenidos a partir de medidas del brillo del cielo nocturno y comparación con modelos estándar, indican que en una ciudad típica argentina de unos 150000 habitantes, el **ABC** resulta ser de 2,5 veces el valor natural esperado a unos 10 km del centro geográfico de la ciudad. El derroche energético inferido a partir de estas medidas, es de aproximadamente un 36 % del gasto en alumbrado público (considerando balastos y siendo mayoritariamente lámparas de mercurio).

Sin embargo, la **CL** encierra problemas que van más allá de lo práctico. Su efecto sobre el medioambiente pone de manifiesto alteraciones a la biota de un lugar (Longcore y Rich, 2004, y referencias allí incluidas) y al ser humano en particular. Existe fuerte evidencia de que las alteraciones del ritmo circadiano humano debido a la exposición nocturna a la luz, pueden estar asociadas con graves enfermedades (Schernhammer et al., 2001; Pauley, 2001). De esta forma, **CL** se ha convertido en una cuestión *medioambiental* seria y aunque este problema no es ajeno a diversos organismos en Argentina (por ejemplo el Departamento de Luminotecnia de la UN-Tucumán, o la Municipalidad de la ciudad de Rosario, provincia de Santa Fe), resulta notable el escasísimo lugar que se le asigna en general en los organismos públicos dedicados a la iluminación y en el ámbito universitario particularmente[1]. Debido a esto, el próximo paso ha sido tender hacia la

- *inserción académica* del tema y su difusión en general.

**CL** es claramente un problema científico-tecnológico y la estrategia más inmediata para proponer la inserción académica resulta ser, justamente, dentro de la **UTN,** universidad dedicada básicamente a la formación de ingenieros, que posee una estructura federal, con más de 30 unidades académicas distribuidas en todo el territorio nacional, que hace posible canalizar proyectos integradores sobre contaminación atmosférica a partir de una red de grupos de trabajo. Dado que la **UTN** es una institución de base tecnológica, se ha centrado el trabajo en los aspectos prácticos para difundir el tema desde lo medioambiental.

## *CL como práctica medioambiental en la Universidad*

Para llevar a cabo esta tarea, se han seleccionado los siguientes puntos que se revelan rápidamente en el estudio de la **CL,** y que se consideran básicos a la hora de presentar un tema novedoso, desde lo práctico como problema medioambiental:

- **CL** *es más fácilmente controlable que la mayoría de las problemáticas ambientales conocidas*: en efecto, no se necesitan grandes despliegues tecnológicos ni procesos anexos para atacar el problema; además, *la mitigación es inmediata y no permanecen efectos residuales.*

- *Su mitigación redunda en beneficios económicos inmediatos o a corto plazo*: debido al control del derroche de energía; por otra parte, este control, no necesariamente genera gastos extra. Lámparas de mercurio pueden irse cambiando por sodio de baja y alta presión según corresponda o sea conveniente. Demás tareas de reemplazo y mejora de luminarias pueden llevarse a cabo dentro de los presupuestos existentes, o generar a corto plazo grandes rentabilidades.

- **CL** incluye entre sus efectos al **ABC**. Este aumento artificial de la luminancia del cielo nocturno, *necesita de la atmósfera para establecerse* (Walker, 1970; Treanor, 1973; Gargstang, 1991). Por lo tanto, en un sentido amplio, la **CL** *necesita de la atmósfera terrestre como agente propagador.*

El simple análisis anterior permite, además, clasificar a la **CL** dentro de las problemáticas de contaminación reconocidas como contaminación atmosférica. Esto es importante a los efectos de no generar nuevas categorías medioambientales y complicar el acceso al tema. Por otra parte, modelos más actuales del **ABC** (Gargstang, 2000) incluyen a los contaminantes atmosféricos en la ecuación de transporte de la radiación incidente, lo cual apoya la clasificación de la **CL** dentro de la polución atmosférica. En particular la contaminación atmosférica se divide actualmente en: contaminación por agentes químicos, biológicos y físicos; por lo tanto, la **CL** debería incluirse en esta última categoría.

En **CL** coordinar esfuerzos es importante, porque sus efectos, en una zona determinada, tienen implicancias más allá de lo local. Por ejemplo, el derroche energético en un sitio perjudica al anillo energético de todo un país; el **ABC** de un lugar puede depender de fuentes ubicadas inclusive a grandes distancias de la zona en cuestión (Cinzano, 2000). El problema está instalado en cada ciudad y un verdadero intento de solución implica necesariamente, una estrategia común.

---

[1] En este sentido, es importante mencionar las iniciativas pioneras de países como Chile, Italia y España.

### *3- Estrategias básicas de presentación de la temática*

Las posibilidades de mitigación anteriormente resaltadas son importantes a los efectos de presentar, ante profesionales de diversas áreas tecnológicas, las estrategias de control de esta polución y que éstas, sean vistas muy accesibles desde el tratamiento práctico. Sin embargo, se ha detectado un problema importante *en la definición o planteo básico de la **CL***: *el tema está fuertemente asociado al ABC*, y debido a esto, las consecuencias de la **CL** suelen reducirse a un segmento de ciudadanos compuesto por astrónomos y observadores aficionados. Esto, además de ser falso, atenta contra la real dimensión del impacto de la **CL** en la sociedad y el medioambiente en general; por lo tanto, como contribución a una estrategia de divulgación en este sentido, se propone el siguiente esquema de división de la problemática, cuyas partes deben tratarse in extenso a la hora de presentar el tema:

1. **Causas,** *¿qué causa la CL?*: iluminación desaprensiva: inadecuada, ineficiente, mal ubicada, etc.

2. **Efectos,** ¿cómo se evidencia?: *demasiada luz en el ambiente*: intrusión lumínica, deslumbramiento, **ABC**, derroche energético.

3. **Consecuencias,** ¿qué problemas trae o qué hipótesis de problemas aparejados a ella pueden plantearse? Estos efectos producen consecuencias para el *conjunto* biota-ser humano de un sitio, lo cual puede aclararse si se propone una diferenciación en :

    - *Consecuencias Culturales*: **CL** produce pérdida del cielo nocturno, esto tiene una profunda implicancia en la sociedad humana en general, no solamente en los observadores profesionales. La historia nos muestra la importancia del cielo en el desarrollo de la humanidad en áreas tan variadas como la ciencia, el arte y la religión; esto es así, más allá del rango de interés de cierto grupo de personas. Por dar un ejemplo práctico, la primera determinación de la velocidad de la luz -y prueba de su finitud- se hizo en base a mediciones astrométricas, doscientos años antes de experiencias en laboratorios.

    - *Consecuencias Económicas*: **CL** produce derroche energético, lo cual se traduce en pérdida de recursos. Argentina consume el 25 % del gasto eléctrico en iluminación; es notable que con tantas crisis energéticas, el tema **CL** no haya tenido una actuación central, aunque se hace hincapié en cuestiones que ya son tenidas en cuenta a la hora de contribuir al ahorro de energía como el *cambio de hora legal*.

    - *Consecuencias Medioambientales*: salud y medioambiente animal y vegetal. Se refiere a cambios en la biodiversidad y a la alteración de los ritmos circadianos (humano, animal y vegetal). Hasta hace poco este punto era el más sujeto a conjeturas, ahora, ya es terreno de análisis fenomenológicos e hipótesis y claros indicios; trabajos publicados, muestran el impacto de la luz en salud (ver por ejemplo, Pauley, 2004 y referencias allí incluidas). En particular la melatonina, es una hormona sumamente importante (aparentemente antioxidante y anticancerígena) que se libera cuando la retina no detecta estímulos luminosos; sin embargo, se ve suprimida ante ciertos niveles de iluminación nocturna (aun con los ojos cerrados), especialmente cuando la luz es de longitudes de onda cercanas al azul.

### *4- Resultados preliminares*

Siguiendo la línea conceptual de análisis descripta en los ítems anteriores, se ha presentado el tema **CL** en el ámbito de la Universidad Tecnológica Nacional, en particular en jornadas en el rectorado de la **UTN**. Esto, ha producido la *inclusión de CL como línea de trabajo en el proyecto **PROIMCA**: Proyecto Integrador para la Mitigación de la Contaminación Atmosférica* de la **UTN**. Actualmente este proyecto, se encuentra en vías de ser reformulado en el rectorado de **UTN**. Se espera, además, dar un puntapié inicial para la inclusión del tema en el ámbito docente universitario. En este sentido, existe material cada vez mejor documentado, en revistas con arbitraje, pero disperso y a veces difícil de obtener debido a la naturaleza interdisciplinaria de la **CL**. El aporte realizado hasta el momento en este área es diverso:

- Preparación de material didáctico (en fase de edición). Aquí se ha notado un problema: falta de uniformidad en las convenciones y definiciones usadas en la medición de la energía radiante. Conceptos como *intensidad*, *flujo* o *luminosidad*, son definidos en forma distinta en el uso estándar, astronómico o en ciencias de la salud; es importante plantear las definiciones desde la física y clarificar los usos y denominaciones particulares.

- Realización de un código abierto basado en Tecnologías Informáticas Libres para modelos y cálculos luminotécnicos relacionados a la **CL** (evaluación del flujo hemisférico superior de una luminaria, estimación de lúmenes por cabeza a partir del **ABC**, etc.). [2]

- Construcción de un fotómetro visual de comparación (ver Fig.3), con fines de divulgación y docencia en el área.

En la faz técnica se han comprado luxómetros para medidas in situ y se espera adquirir un telescopio reflector de 20 cm de abertura con cámara CCD para procesamiento digital de imágenes del cielo y demás periféricos.

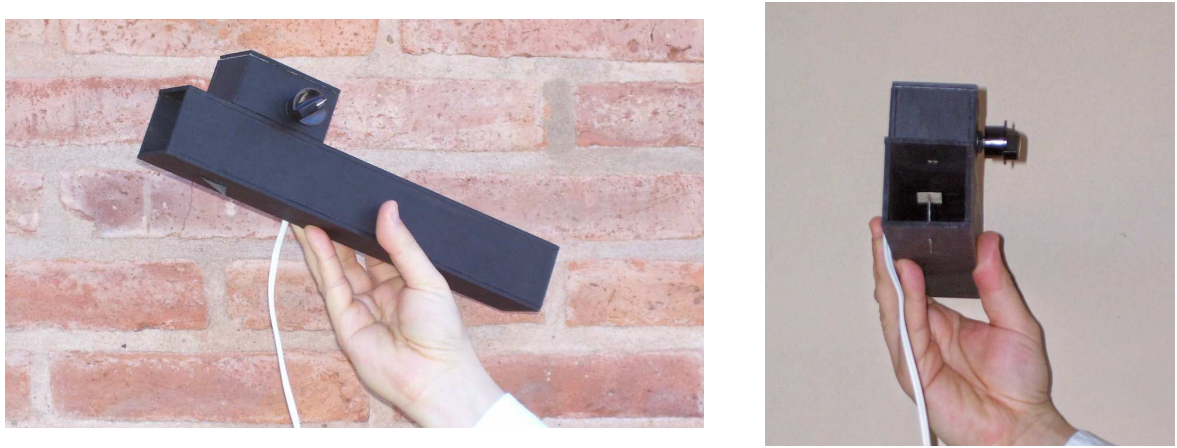

*Fig. 3 Fotómetro visual de comparación desarrollado en el GEA UTN-FRSN. La perilla regula la intensidad de una lamparita de 5V, a través de un potenciómetro; esto produce la iluminación de una pantalla pequeña de comparación, lo cual permite la estimación de la luminancia del cielo en unidades instrumentales. Un dispositivo similar ha sido propuesto por Pike y Berry (1978).*

## *5-Conclusiones y perspectivas*

En este trabajo se argumenta a favor de la inclusión de la **CL** dentro del área contaminación física de la atmósfera y su inserción en proyectos integradores en ámbitos académicos-científicos de la República Argentina. El tema, presentado según los lineamientos aquí descriptos, ha recibido la atención y el apoyo de los coordinadores del proyecto **PROIMCA** de la **UTN**, y se encuentra en su etapa de implantación. Las actividades propuestas para **CL** dentro de este proyecto, son:

- Formar recursos humanos en esta temática para profesionales de *origen o áreas diversas*.

---

[2] El material enumerado estará en breve disponible a través de la página WEB de la UTN-FRSN: www.frsn.utn.edu.ar.

- Organizar grupos de trabajo comunales o municipales, con objeto de plantear estrategias *zonales de mitigación*: este es el tema central. Los tipos de luces y luminarias, las inversiones iniciales en iluminación y sus posibilidades de recuperación por ahorro son muy dependientes de las instalaciones particulares de un sitio. Estas estrategias deben contemplar especialmente: división en zonas de interés medioambiental e iluminación según clases de vía (alumbrado vial).

- Realizar relevamiento y diagnóstico de los niveles de **CL** a nivel nacional.

- Generar ámbitos científico-académicos de discusión y estudio de la **CL** que contemplen la realización de cursos de capacitación y presentación de trabajos específicos: incluir a la **CL** en las reuniones científicas de contaminación atmosférica.

- Generar una base de datos a nivel nacional en el área.

- Trabajar sobre normativas de acuerdo a las necesidades detectadas; aquí hay varias fuentes que merecen citarse: publicaciones de la CIE (CIE-126, 1997) en lo referente a zonas de importancia medioambiental, toques de queda, etc., y normativas existentes como las generadas en la *Oficina Técnica para la Protección de la Calidad del Cielo* en Canarias, España (OTPC, 2001); la Comisión Nacional para la Protección del Medioambiente en Chile (CONAMA, 1997) y la producida por Puig, San Martín y Torra (2001) en Cataluña, España.

- Generar programas de divulgación, servicio y transferencia al medio.

Aunque la iluminación es necesaria y la actividad humana conlleva efectos no deseados, estos deben minimizarse o reducirse en lo posible. Respecto a la difusión y concientización, se ha detectado que, en general, el exceso de luz o la mala iluminación parece ser un problema menor. La **CL** deja de ser un problema menor cuando es clasificada y presentada adecuadamente en causas-efectos-consecuencias, esto expone el problema en su real dimensión y no lo relega a un grupo de ciudadanos. Por otra parte, cuando se discute lo relativamente sencillo de mitigar el problema (instalar luminarias y lámparas adecuadas, bien orientadas y utilizadas con criterio razonable) y lo beneficioso económicamente que resulta atender a estos estándares, rápidamente surge la toma de conciencia y la posibilidad de resolver el tema. Estas cuestiones, lejos de poner a la iluminación como un problema, la jerarquizan, haciendo de ella un sujeto científico-tecnológico complejo y apasionante.

La **CL** es un problema medioambiental zonal integrado en todo un país. El indicador de diagnóstico es el aumento del brillo del cielo nocturno, éste puede modelarse fácilmente y a partir de allí, es posible estimar, por ejemplo, la emisión de flujo superior de cierta población. Sin embargo, el medioambiente puede verse afectado por un flujo con elevación menor al plano horizontal; es necesario profundizar en la mitigación de la *intrusión lumínica* y ampliar el concepto de tal forma de tener en cuenta a la modificación de los niveles naturales de iluminación de sitios naturales, en especial, lugares de utilidad reproductiva para ciertas especies animales.

Aunque a escala mundial la mayor responsabilidad por flujo de emisión superior la tienen los países del Norte (los más desarrollados), América Latina como reserva natural (en medioambiente y cielos: los telescopios más importantes se están colocando en América del Sur), no debe perder la oportunidad además, de optimizar el gasto en energía y ser ejemplo en manejo de recursos medioambientales.

## 6- Referencias